\documentclass[aps,superscriptaddress,nofootinbib,showpacs,eqsecnum,preprint,tightenlines]{revtex4}
\usepackage{hyperref}
\usepackage{epsfig,rotating}
\usepackage{amsmath,amssymb}
\usepackage{dsfont}
\usepackage{simplewick}

\newcommand{\be}{\begin{equation}}
\newcommand{\ee}{\end{equation}}
\newcommand{\bea}{\begin{eqnarray}}
\newcommand{\eea}{\end{eqnarray}}
\newcommand{\nn}{\nonumber}
\newcommand{\vx}{\vec{x}}

\newcommand{\vp}{\vec{p}}
\newcommand{\vP}{\vec{P}}
\newcommand{\vq}{\vec{q}}
\newcommand{\vQ}{\vec{Q}}
\newcommand{\vk}{\vec{k}}

\newcommand{\oO}{\overline{\Omega}}
\begin{document}

\title{Dynamics of disentanglement, density matrix and coherence in neutrino oscillations.  }
\author{Jun Wu}\email{juw31@pitt.edu}  \affiliation{Department of Physics and
Astronomy, University of Pittsburgh, Pittsburgh, PA 15260}
\author{Jimmy A. Hutasoit} \email{jhutasoi@andrew.cmu.edu}
\affiliation{Department of Physics, Carnegie Mellon University,
Pittsburgh, PA 15213, USA}
\author{Daniel Boyanovsky}
\email{boyan@pitt.edu} \affiliation{Department of Physics and
Astronomy, University of Pittsburgh, Pittsburgh, PA 15260}
\author{Richard Holman}
\email{rh4a@andrew.cmu.edu} \affiliation{Department of Physics, Carnegie Mellon University, Pittsburgh, PA 15213, USA}

\date{\today}

\begin{abstract}
In charged current weak interaction processes,  neutrinos are
  produced in an entangled state with the charged lepton. This correlated state is
disentangled by the   measurement of the charged lepton in a detector at the production site. We study the dynamical aspects of disentanglement,
propagation and detection, in particular the conditions under which
the disentangled state is a coherent superposition of mass
eigenstates.   The appearance and disappearance far-detection
processes are described from the time evolution of this disentangled
``collapsed'' state. The familiar quantum mechanical interpretation
and factorization of the detection rate emerges when the quantum
state is disentangled on time scales \emph{much shorter} than the inverse
oscillation frequency, in which case the final detection rate factorizes
in terms of the usual quantum mechanical transition probability
provided the final density of states is insensitive to the neutrino
energy difference. We suggest \emph{possible} corrections for short-baseline experiments. If the charged lepton is unobserved, neutrino
oscillations and coherence are described in terms of a reduced
density matrix obtained by tracing out an un-observed charged
lepton. The diagonal elements in the mass basis describe the production of mass eigenstates whereas the off diagonal ones provide a measure of coherence. It is shown that coherences are of the same order of the diagonal terms on time scales up to the inverse
oscillation frequency, beyond which the coherences oscillate as a result of the interference between mass eigenstates.

\end{abstract}


\maketitle

\section{Introduction}

Neutrino masses, mixing and oscillations are the clearest evidence yet of physics beyond the standard model \cite{book1,book2,book3}. They provide an explanation of the solar neutrino problem \cite{msw,book4,haxtonsolar} and have other important phenomenological \cite{book1,book2,book3,grimuslec,kayserlec,mohapatra,degouvea}, astrophysical \cite{book4,book5,haxton} and cosmological \cite{dolgovcosmo} consequences. Furthermore, they offer an incredible example of macroscopic quantum coherence, where in long baseline oscillation experiments this coherence is maintained over hundreds of kilometers.

Given the importance of neutrino oscillations in pointing us to the next layer beyond that of the standard model, it is not unreasonable to make sure we understand the domain of validity of the various calculations of oscillation probabilities. In particular, the standard approach of treating neutrino oscillations by analogy with Rabi-oscillations in a two state system (see for
example, \cite{book1,book2,book3,kayserlec,grimuslec} and references therein) while simple and intuitive, is not necessarily the last word on the subject. Deeper investigations of this basic paradigm have already raised a number of important and fundamental questions \cite{kayser,rich,nauenberg,lipkin1} that are still being debated \cite{lipkinuncertainty,lipkingsi,aksmir,akmerev}.

If we truly want to understand the results from oscillation experiments, we need to factor in both the production and detection mechanisms. In term of its detection, the neutrino is never directly detected. We only know that it is there due to the charged leptons associated with its weak interaction at the detector. As for the production mechanism,  the neutrino state is produced by the decay of a parent particle via charged current interactions. Interestingly, a discussion of coherence aspects of neutrinos \cite{giunticohe,stock,glashow,dolgov,losecco,hamish} has recognized that the neutrino state produced by the decay of a parent particle via charged current interactions is in fact \emph{entangled} with that of the charged lepton.


How does this entangled state evolve? As usual, there are two choices. If we measure the charged lepton, this disentangles the quantum state at a time determined  either by the detection process of the charged lepton at the source or by  stopping at a ``beam dump'' near the source.  This disentangled neutrino state then evolves further in time and is detected at the  ``far detector'' via another charged lepton. Therefore the production, disentanglement and detection involve \emph{two times}: the time at which the charged lepton is observed at the source (if observed at all) or stopped at a beam dump near the source,  which is when the neutrino/charged lepton state is disentangled, and the time when the neutrino is observed via a charged lepton in the far detector. We note that for long-baseline experiments these two time scales are widely separated.

If the charged lepton is not observed,   the state is described by a reduced \emph{density matrix}, which is obtained by tracing out the charged lepton. In this case, we can track the behavior of the off-diagonal elements, \emph{i.e.}, the coherences which contain information on the interference between the different mass eigenstates.

What we do in this article is to use a full field theoretic approach to follow this time evolution explicitly. We start from a simplified model that encodes the salient interactions necessary to analyze the development of oscillations. Using this model, we calculate the form of the state produced when the parent particle (be it a $W$ boson or a pion) decays into the flavor neutrino and its associated charged lepton. We then use the time evolution operator, calculated  within perturbation theory from the interaction Lagrangian, to evolve this state until it reaches the detector. At this point, we obtain the overlap with the final states observed at the detector.

Our main observation is that oscillation experiments are {\em two-time} experiments, the two times being the time at which the original charged lepton is measured and that when the final one is measured. While this does not seem earth-shattering, what is extremely interesting is the interplay between these times (and/or time differences) and other time scales in the problem such as the oscillation time scale and the time scale upon which it can be said that the two mass eigenstates have been resolved. In fact, one of our results is that the treatment of the flavor eigenstate as the usual coherent mixture of the two mass eigenstates requires that the original entangled state already be disentangled long before an oscillation time. While this is true for long baseline experiments, it is not necessarily so for short baseline ones, and this could give rise to a reinterpretation of their results to take the dependence on the disentangling time into account.

In all of this, we would like to emphasize that we are just following the tenets of quantum mechanics in our calculation: prepare a state, construct and then use the relevant time evolution operator and lastly, make a measurement on that state by overlapping it with various states of interest at the later time. Following the time evolution in detail as we do gives us the access to finer details of the behavior of the state than what would be obtained when one uses Fermi's Golden Rule, which requires taking an infinite time limit, to directly calculate rates and the like. In our formalism, the finite-time effects become explicit and this allows us to fully understand the domain of validity of the various results in the literature.

\section{A   Model of ``Neutrino'' Oscillations}\label{sec:model}

 Let us start by introducing our model. In order to exhibit the main results in a clear and
simple manner, we introduce a bosonic model that describes mixing,
oscillations and charged current weak interactions reliably without
the complications associated with fermionic and gauge fields. We can do so because the technical complications associated
with spinors and gauge fields are irrelevant to the
physics of mixing and oscillations, as is obviously manifest in
meson mixing. Our model is
defined by the following Lagrangian density\be \mathcal{L} =
\mathcal{L}_0[W,l_\alpha]+ \mathcal{L}_0[\nu_\alpha] +
\mathcal{L}_{\rm int}[W,\l_{\alpha},\nu_\alpha] ~~;~~ \alpha=e,\mu
\label{totallag} \ee with \be {\cal L}_0[\nu] =
\frac{1}{2}\left[\partial_{\mu}\Psi^T
\partial^{\mu}\Psi  -\Psi^{T} \mathbb{M}\Psi  \right] \label{nulag}\, ,
\ee where  $\Psi$ is a flavor doublet representing the neutrinos\be
\Psi = \left(
             \begin{array}{c}
               \nu_e \\
               \nu_\mu \\
             \end{array}
           \right), \label{doublet}\ee and $\mathbb{M}$ is the mass matrix
            \be \mathbb{M} = \left(
                               \begin{array}{cc}
                                 m_{ee} & m_{e\mu} \\
                                 m_{e\mu} & m_{\mu \mu} \\
                               \end{array}
                             \right)\,. \label{massmtx}\ee
The interaction Lagrangian is similar to the charged current interaction of the standard model, namely
\be {\cal L}_{\rm int} (\vx,t) = g\, W(\vx,t)\Big[ l_{e}( \vec{x},t)\,\nu_{e}(
\vec{x},t)+ l_{\mu}( \vec{x},t)\,\nu_{\mu}( \vec{x},t)\Big],
\label{Interaction} \ee where $g$ is the coupling constant. $W(x)$
represents the vector boson, or alternatively the pion field, and
$l_{\alpha}$, $\alpha = e,\mu$ the two charged leptons. The mass
matrix is diagonalized by a unitary transformation \be
U^{-1}(\theta) \, \mathbb{M}\, U(\theta) = \left(
                                                           \begin{array}{cc}
                                                             m_1 & 0 \\
                                                             0 & m_2 \\\end{array}\right)
 ~~;~~ U(\theta) = \left(                                                                                  \begin{array}{cc}                                                                                       \cos \theta & \sin \theta \\                                                                                       -\sin\theta & \cos\theta \\                                                                                     \end{array}                                                                                   \right) . \label{massU}\ee In terms of the
  doublet of mass eigenstates,  the flavor doublet can be expressed as
  \be \left(
                                     \begin{array}{c}
                                       \nu_e \\
                                       \nu_\mu \\
                                     \end{array}
                                   \right) = U(\theta)\,\left(
                                               \begin{array}{c}
                                                 \nu_1 \\
                                                 \nu_2 \\
                                               \end{array}
                                             \right) \,.\label{masseigen}\ee
This bosonic model clearly describes charged current weak
interactions reliably as it includes all the relevant aspects of
mixing and oscillations.

We consider ``neutrino'' oscillation experiments following the
interaction processes illustrated in Fig.~\ref{fig:experiment
setup}, \bea W\rightarrow l_{\alpha}+\nu_{\alpha} \rightsquigarrow &
\left\{\begin{array}{ll} W+l_{\beta}~~,\beta \neq \alpha & \mbox{
appearance  } \\ W+l_{\alpha} & \mbox{ disappearance
 } \end{array}\right.\,. \eea
 \begin{figure}[h]
\begin{center}
\includegraphics[width=7cm,keepaspectratio=true]{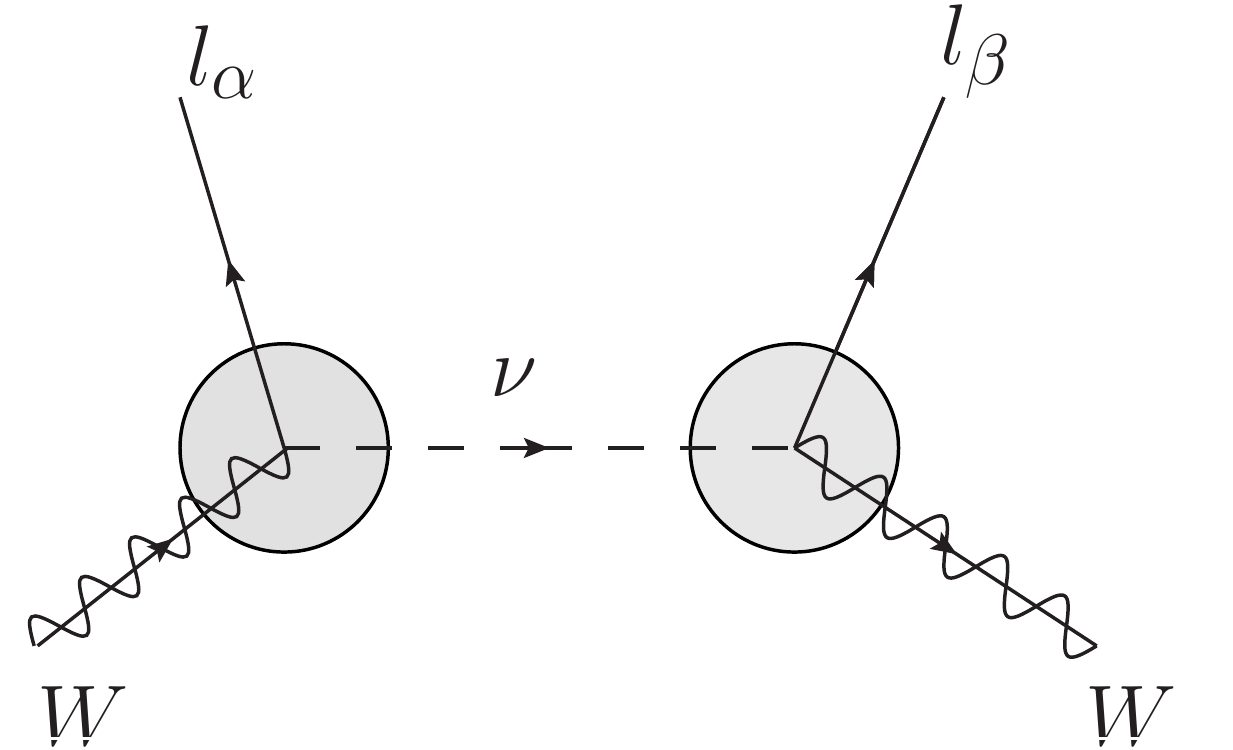}
\caption{Typical experiment in which the charged leptons are measured at the source and far
detector and the neutrino is an intermediate state.}
\label{fig:experiment setup}
\end{center}
\end{figure}

Throughout this work, we will use plane wave states for simplicity of exposition, though we discuss some of the modifications that the use of wave packets would require later in this section.

      In order to study aspects of coherence we consider a simplified
      interaction Lagrangian density \be \mathcal{L}_I = g\, W \,e
      \,\nu_e = g \,W \,e (\cos \theta\, \nu_1 +\sin \theta \, \nu_2),
      \label{lagsim}\ee focusing only on one lepton, which we refer
      to as the ``electron.'' The full
      coupling as in Eq. (\ref{Interaction}) can be treated similarly
      without modifying the main conclusions. Although $W$ may be interpreted as a charged vector boson,
      the analysis is obviously the same if it describes a pion
      field.

      We can study aspects of coherence by focusing on the Fock
      state obtained upon evolution of the decaying initial state.
    We consider a plane-wave Fock initial state $\big|W(\vk)\big\rangle$ at $t_i=0$.   The time evolution operator is \be e^{-iH(t-t_i)} = e^{-iH_0t}~U(t,t_i)~e^{iH_0t_i}, \label{timeoper}\ee  where \be U(t,t_i) = T~e^{i\int_{t_i}^t ~d^3x\,dt\,\mathcal{L}_{\rm int}(\vec{x},t)} \label{Uevol}\ee is the time evolution operator in the interaction picture and $H_0$ is the non-interacting Hamiltonian.

    By time $t$, the initial state has
    evolved into  $\big|\Psi(t)\rangle = \big|W(\vk)\big\rangle ~e^{-iE^W_{\vk}t}+ \big|\Psi_e(t)\big\rangle$. To lowest order in the interaction, we find the second term to be \be  \big|\Psi_e(t)\big\rangle =  i g~e^{-iH_0t}~
    \int_0^t dt'\int d^3x ~\Big[W(\vec{x},t')\, e(\vec{x},t') (\cos \theta\, \nu_1(\vec{x},t')+ \sin \theta\,
    \nu_2(\vec{x},t'))\Big]~\big|W(\vk)\big\rangle, \label{newstate} \ee
    where all the fields are in the interaction picture. Though the  field
    operator $W$ can either annihilate the initial state or create another $W$ particle,  the state with two $W$ particles
    features faster oscillations that will average out. In what follows, we consider only the Fock state resulting
    from the annihilation of the initial particle, leading to the state
\bea  \big|\Psi_e(t)\big\rangle & \simeq &
 \frac{g}{2\sqrt{2VE^W_{\vk}}}~e^{-iE^W_{\vk}t}~
    \sum_{\vq}    \Bigg\{
     \frac{\sin\theta}{\sqrt{\Omega_{2, \vp} ~ E^{e}_{\vq}}}~\Big|\nu_{2,\vp}\rangle \Big|e_{\vq} \rangle
    \Bigg[\frac{e^{ i(E^W_{\vk}-E^{e}_{\vq}-\Omega_{2,\vp} )t}-1}{
    (E^W_{\vk}-E^e_{\vq}-\Omega_{2,\vp} )} \Bigg]
    \nonumber \\ & + & \frac{\cos\theta}{\sqrt{ \Omega_{1,\vp}~
     E^e_{\vq}}}~\Big|\nu_{1,\vp}\rangle \Big|e_{\vq} \rangle
    \Bigg[\frac{e^{ i(E^W_{\vk}-E^e_{\vq}-\Omega_{1,\vp} )t}-1}{ (E^W_{\vk}-E^e_{\vq}-
    \Omega_{1,\vp} )} \Bigg]
     \Bigg\}~~;~~ \vp = \vk-\vq \label{finstate}  \eea in which the
   electron and the neutrinos are \emph{entangled}\footnote{The result for the wavefunction in Ref. \cite{glashow} may be understood using a (non-perturbative) Wigner-Weisskopf approximation for the decaying parent particle, replacing $E_W \rightarrow E_W-i\Gamma_W$. Taking $t \gg 1/\Gamma_W$ in the integral replaces the brackets in (\ref{finstate}) by $1/(E_W-E^e-\Omega_j -i\Gamma_W)$ whose absolute value is proportional to $\delta(E_W-E^e-\Omega_j )/\Gamma_W$.}. The neutrino state that is entangled with the
    muon is obtained from (\ref{finstate}) by replacing $\cos \theta \rightarrow -\sin \theta~;~ \sin\theta \rightarrow  \cos\theta$.

    In what follows, we consider ultrarelativistic neutrinos and write
   \be \Omega_1 = \overline{\Omega} - \Delta~~;~~
  \Omega_2 = \overline{\Omega} + \Delta, \label{URO}\ee where
   \be   \overline{\Omega}  = \left[ p^2 +\frac{m^2_1+m^2_2}{2}\right]^\frac{1}{2} ~~;~~ \Delta= \frac{\delta m^2}{4\overline{\Omega}  }
   ~~;~~\delta m^2 = m^2_2-m^2_1,  \label{diffs}\ee taking
   $\Delta \ll \oO$ as is the case for ultrarelativistic nearly degenerate neutrinos.

\subsection{Unobserved daughter particles: time evolution of the
density matrix}

     If the electrons (or
    daughter particle in Ref. \cite{glashow}) are not observed, they can be
    traced out of the \emph{density matrix} $ \big|\Psi_e(t)\big\rangle \big \langle
    \Psi_e(t)\big|$. This gives the \emph{reduced} density matrix \bea
    \rho_r(t)& =& \mathrm{Tr}_{e}\big|\Psi_e(t)\big\rangle \big \langle
    \Psi_e(t)\big| \nonumber \\
    & = &   \frac{g^2}{8 V E^W_{\vk}} \sum_{\vq}    \Bigg\{
    \frac{\sin^2\theta}{\Omega_{2, \vp} ~ E^e_{\vq}} \, \Big|\nu_{2,\vp}\rangle \langle
    \nu_{2,\vp}\Big| ~
    \Bigg[\frac{\sin\Big(\big(E^W_{\vk}-E^e_{\vq}-\Omega_{2,\vp}  \big)\frac{t}{2}\Big)}{\big(E^W_{\vk}-E^e_{\vq}-\Omega_{2,\vp}\big)/2 }
     \Bigg]^2 \nonumber \\ && + \,  \frac{\cos^2\theta}{\Omega_{1, \vp} ~ E^e_{\vq}}\,\Big|\nu_{1,\vp}\rangle \langle
    \nu_{1,\vp}\Big| ~
    \Bigg[\frac{\sin\Big(\big(E^W_{\vk}-E^e_{\vq}-\Omega_{1,\vp}  \big)\frac{t}{2}\Big)}{\big(E^W_{\vk}-E^e_{\vq}-\Omega_{1,\vp}\big)/2 }
     \Bigg]^2 \nonumber \\ & &+ \, \frac{\sin2\theta}{2E^e_{\vq}\, \sqrt{\Omega_{2, \vp}~\Omega_{1,
     \vp}}}\Bigg[\frac{\sin\Big(\big(E^W_{\vk}-E^e_{\vq}-\Omega_{2,\vp}  \big)\frac{t}{2}\Big)}{\big(E^W_{\vk}-E^e_{\vq}-\Omega_{2,\vp}\big)/2 }
     \Bigg]\Bigg[\frac{\sin\Big(\big(E^W_{\vk}-E^e_{\vq}-\Omega_{1,\vp}  \big)\frac{t}{2}\Big)}{\big(E^W_{\vk}-E^e_{\vq}-\Omega_{1,\vp}\big)/2 }
     \Bigg]\nonumber \\ & &\times \, \Bigg[ e^{ - i \frac{\delta m^2}{4\oO} t} \, \Big|\nu_{2,\vp}\rangle \langle
    \nu_{1,\vp}\Big|+ e^{  i \frac{\delta m^2}{4 \oO} t} \, \Big|\nu_{1,\vp}\rangle \langle
    \nu_{2,\vp}\Big|\Bigg] \Bigg\}~~;~~\vp = \vk-\vq\,.
    \label{redrho}\eea
This expression contains remarkable information. The function
$\sin^2(xt)/x^2$ is the usual ``diffraction'' function of Fermi's
Golden rule. In the formal long time limit $\sin^2(xt)/x^2
\rightarrow \pi\,t\,\delta(x)$, the first two terms of the density
matrix, which are the diagonal entries in the mass basis, describe the production process
of the mass eigenstates. As will be seen below, in the long time
limit, the time derivative of these two terms yields the production
rate  for the mass eigenstates. In the formal $t\rightarrow \infty$ limit, these
are the diagonal terms obtained in Ref. \cite{glashow}, where in that
reference, the product of delta functions is again understood as the
total time elapsed times an energy conserving delta function.

\begin{figure}[h!]
\begin{center}
\includegraphics[height=3in,width=3in,keepaspectratio=true]{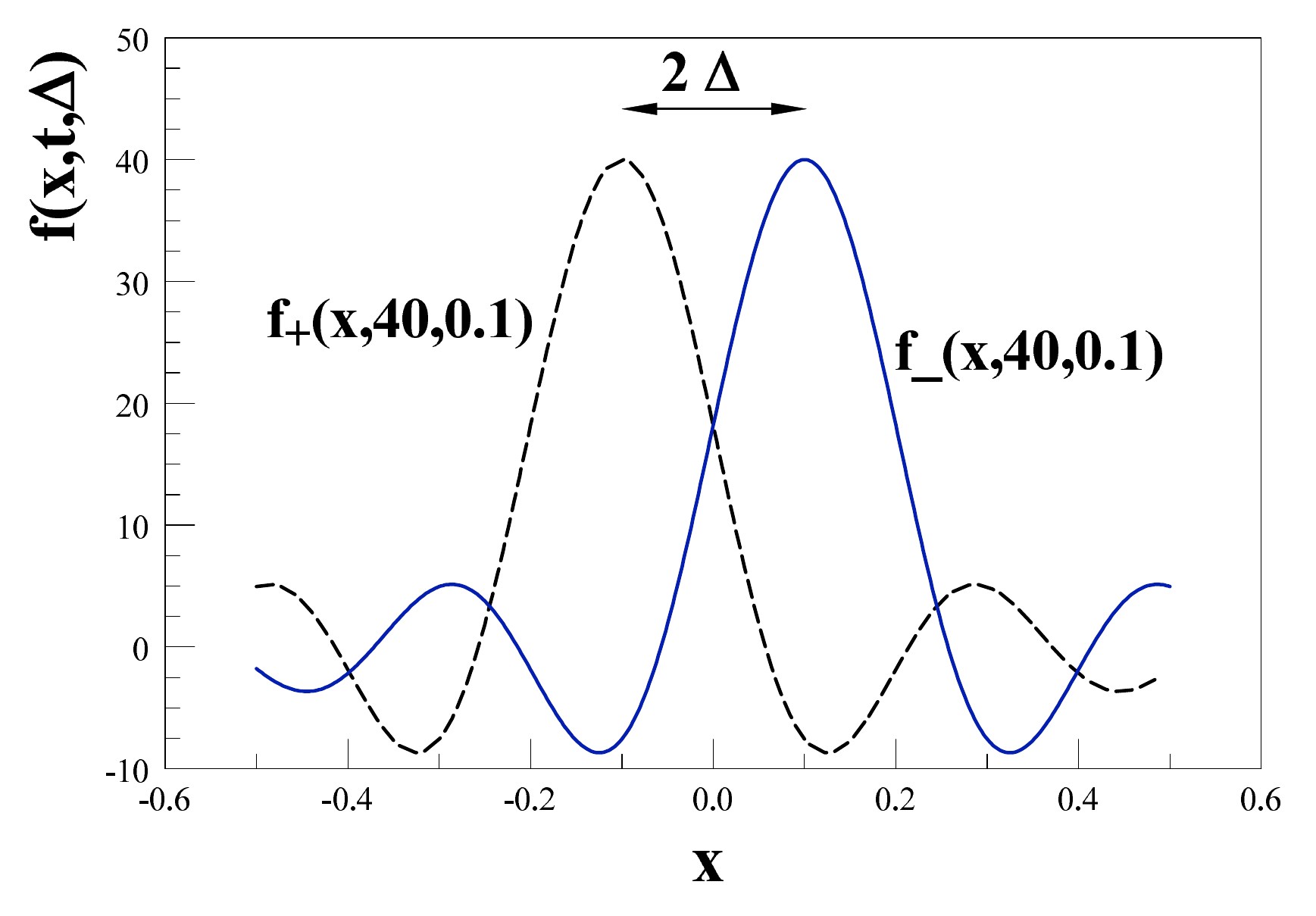}
\includegraphics[height=3in,width=3in,keepaspectratio=true]{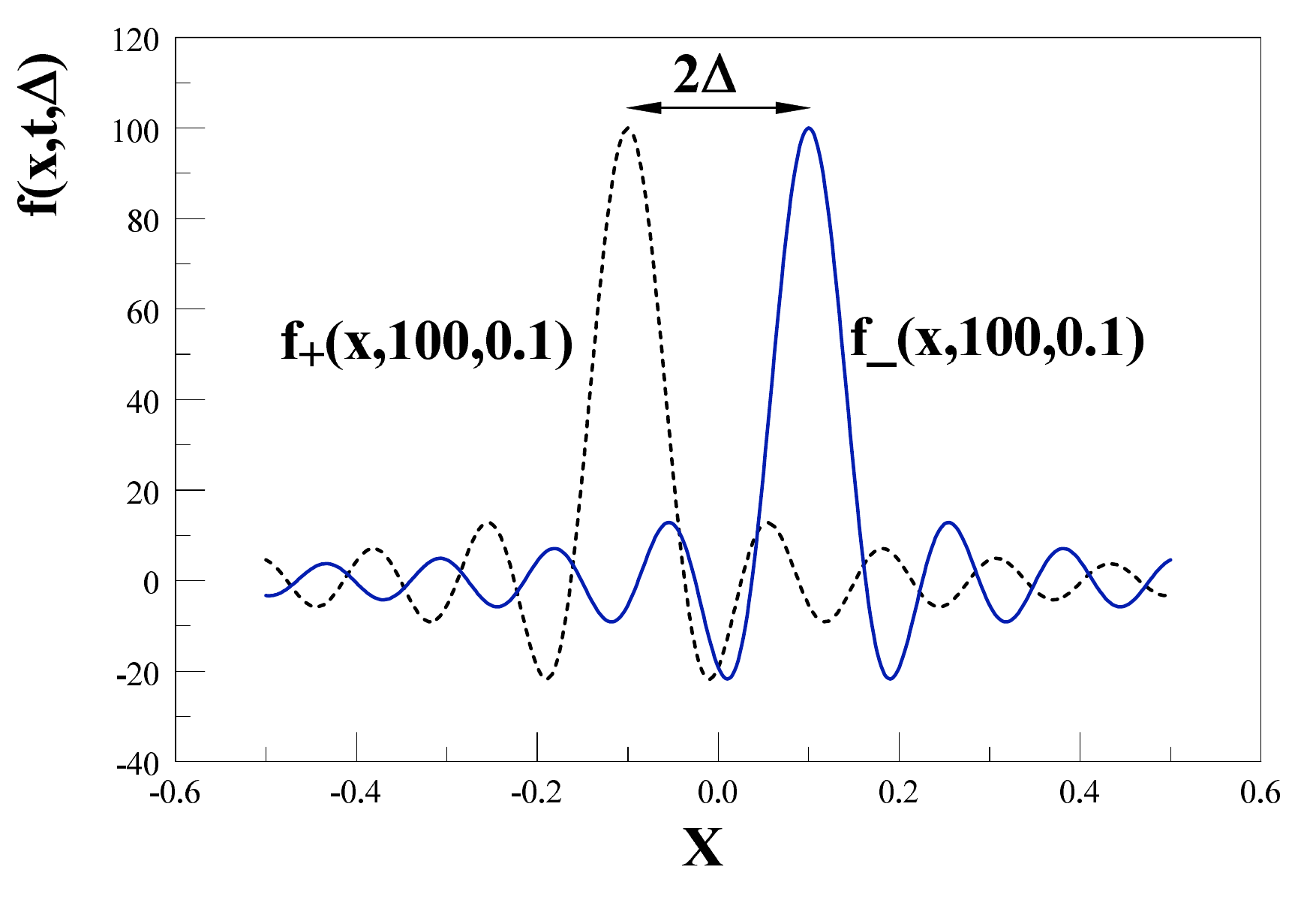}
\caption{The functions $f_\pm(x,t,\Delta)$ vs. $x$ for
$t=40,100$, $\Delta=0.1$} \label{fig:fpm}
\end{center}
\end{figure}

\begin{figure}[h!]
\begin{center}
\includegraphics[height=3in,width=3in,keepaspectratio=true]{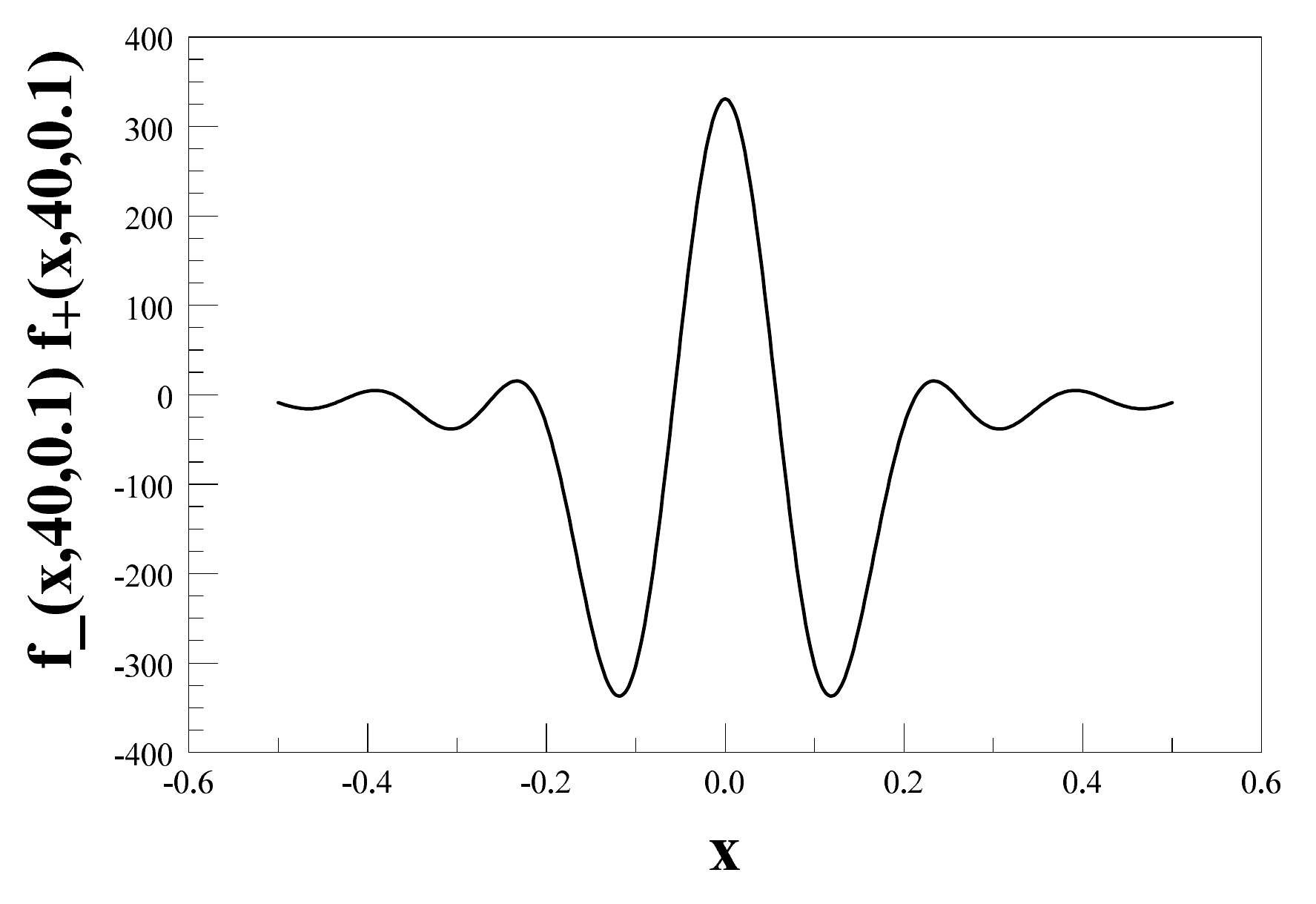}
\includegraphics[height=3in,width=3in,keepaspectratio=true]{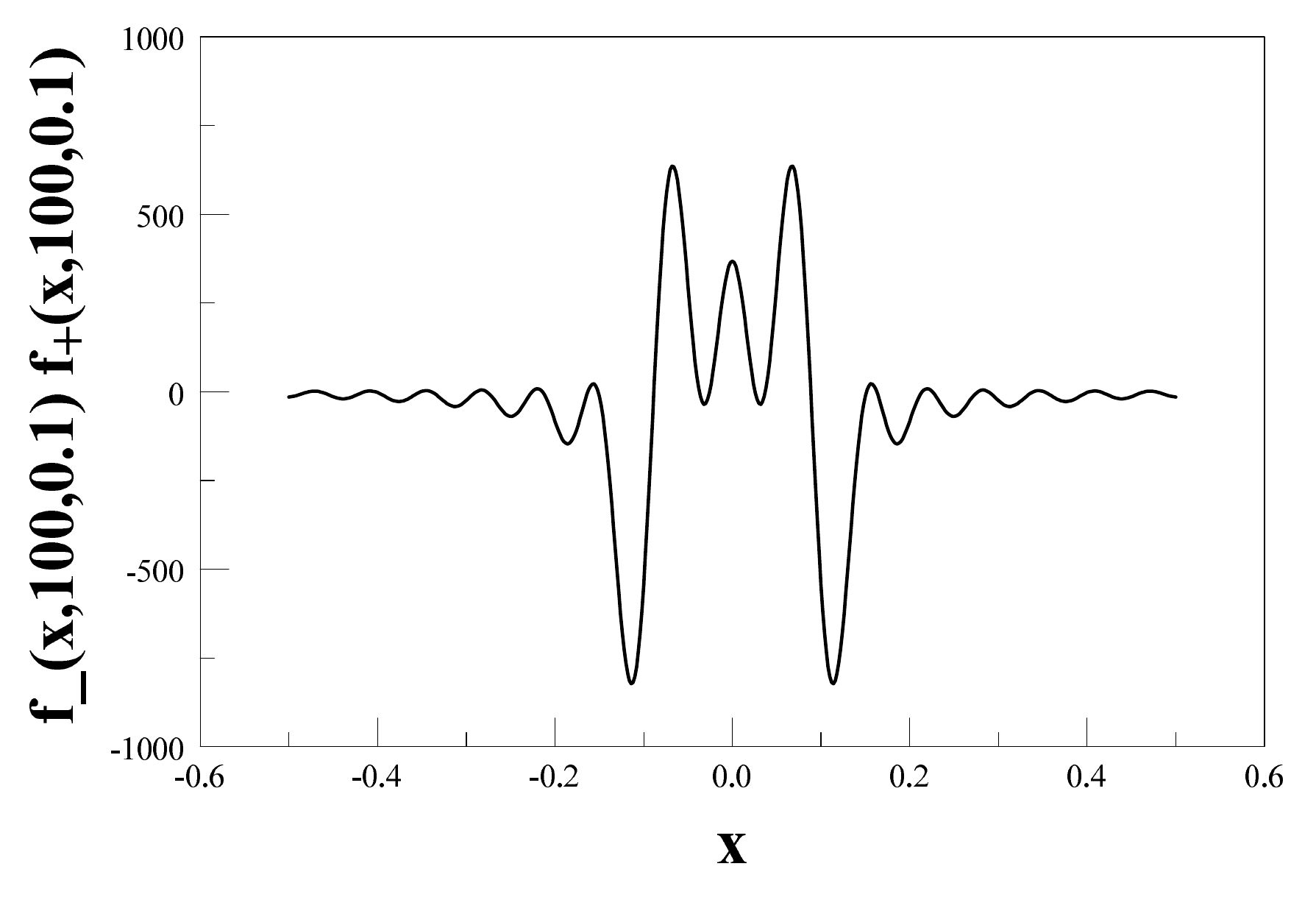}
\caption{The products $f_-(x,t,\Delta)f_+(x,t,\Delta)$ vs. $x$ for
$t=40,100$,$\Delta=0.1$} \label{fig:fprods}
\end{center}
\end{figure}

The off-diagonal terms in the last line of (\ref{redrho}) describe
the ``coherences'' and display the oscillatory phases from the
interference of the mass eigenstates. The time dependent factors
of the off-diagonal density matrix elements determine the
\emph{coherence} between the mass eigenstates and will be a
ubiquitous contribution in the real time description of oscillations
that follows below. The functions \be f_{\pm}(x,t;\Delta) = \frac{2
\sin\Big[\big(x\pm \Delta\big)\frac{t}{2}\Big]}{\big(x\pm
\Delta\big)} ~~;~~ x=E^W_{\vk}-E^e_{\vq}-\oO_{\vp} ~~;~~ \Delta =
\frac{\delta m^2}{4\overline{\Omega}}\,,  \label{fpm}\ee   are strongly peaked at $x\pm \Delta=0$ with height $t$ and
width $\sim 2\pi/t$. In the infinite time limit, $f_\pm(x,t,\Delta)
\rightarrow 2\pi \,\delta(x\pm\Delta)$ and therefore their product would
vanish in this limit, leading to the
vanishing of the coherence. This is the result obtained in
Ref. \cite{glashow}. However, \emph{at finite time }$t$, they feature
a non-vanishing overlap when $2\Delta \lesssim 2\pi/ t$. We
recognize this   as the condition for oscillations. We note
that $t \sim \pi/\Delta$ yields a \emph{macroscopically} large
time scale. The functions $f_\pm(x,t,\Delta)$ and their products are
depicted in Figs. (\ref{fig:fpm},\ref{fig:fprods}) for the values
$\Delta=0.1$, $t=40,100$, respectively.

It is straightforward to
find \bea f_-(x,t,\Delta)f_+(x,t,\Delta) & =  & \frac{\sin(\Delta\,t)}{\Delta}\Bigg[\frac{\sin[(x-\Delta)t]}{(x-\Delta)}+
 \frac{\sin[(x+\Delta)t]}{(x+\Delta)}\Bigg]   \nonumber\\ & & + \, 2\frac{\cos(\Delta\,t)}{\Delta}\Bigg[\frac{\sin^2[(x-\Delta)\frac{t}{2}]}{(x-\Delta)}- \frac{\sin^2[(x+\Delta)\frac{t}{2}]}{(x+\Delta)}\Bigg]
\label{formula} \,.\eea In the long time limit, the terms in the
first bracket yield a sum of delta functions at $x = \pm \Delta$.
Upon integrating the product of $f_-f_+$ with functions of compact
support, the contribution from the second line in (\ref{formula}) is
negligible in the long time limit.
 Therefore, the long time limit of $f_+f_-$ can be
 replaced by
   \be f_-(x,t,\Delta)f_+(x,t,\Delta)= \pi \, \frac{\sin(\Delta\,t)}{\Delta}\Bigg[\delta(x-\Delta) + \delta(x+\Delta) \Bigg]. \label{ltff}\ee
During a time interval  $t\ll 2\pi/\Delta$,   the product
$f_-(x,t,\Delta)f_+(x,t,\Delta)\sim \pi\, t [\delta(x-\Delta) +
\delta(x+\Delta)] $ grows linearly in time until it begins to oscillate with frequency $2\pi/\Delta$ for  $t > 2\pi /\Delta$.

Therefore, we conclude that upon integration with a smooth density of
states, the off-diagonal terms in the density matrix grow linearly in
time for $t \ll t_{osc} = 2\pi/\Delta$, but feature a bound
oscillatory behavior of frequency $\Delta$ for $t \gtrsim
2\pi/\Delta$.

\begin{figure}[h!]
\begin{center}
\includegraphics[width=4in,keepaspectratio=true]{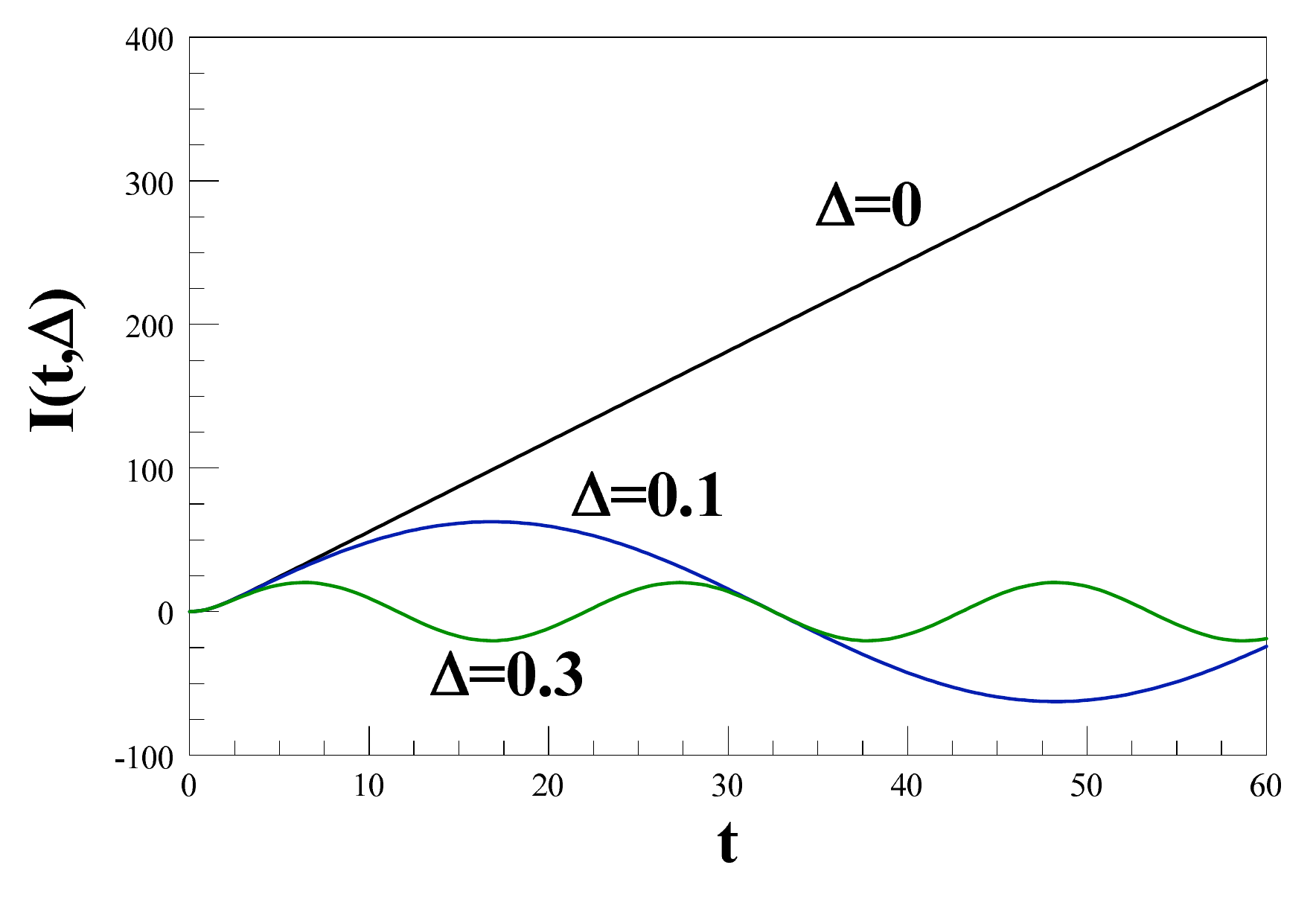}
\caption{The integral $I(t,\Delta) = \int^\infty_{-\infty} e^{-x^2}
f_+(x,t,\Delta)f_-(x,t,\Delta) dx $ vs. $t$ for
$\Delta=0,0.1,0.3$ } \label{fig:integral}
\end{center}
\end{figure}

The diagonal terms,  {\it i.e.}, the  first two terms in the reduced density
matrix (\ref{redrho}), are proportional to $4\sin^2(x\frac{t}{2})/x^2
\rightarrow 2\pi ~t \delta(x)$. This shows that the coherences or off diagonal terms
are linear in time and of the \emph{same order} as the diagonal
elements for $t \lesssim  t_{osc} = 2\pi/\Delta$, but are of
$\mathcal{O}(1/\Delta t)$ and oscillate   compared to the
diagonal terms for $t \gg t_{osc}$.    This behavior is displayed in Fig. (\ref{fig:integral}),
where as an example we consider a smooth density of states and   the
integral \be I(t,\Delta) = \int^\infty_{-\infty} e^{-x^2}
f_+(x,t,\Delta)f_-(x,t,\Delta) dx. \label{testint}\ee
The case
$\Delta =0$ describes either of the diagonal terms and is linearly
secular in time. This figure clearly shows the slow oscillations for
$t \gtrsim 1/\Delta$.

Therefore, the approximation (\ref{ltff}) is reliable in the long
time limit and upon  integration  with functions of compact support. We see that for large $t$, but $t\Delta \ll 1$,  $
f_+(x,t,\Delta)f_-(x,t,\Delta) \rightarrow \pi t
\big[\delta(x-\Delta)+\delta(x+\Delta)\big]$ and for $\Delta
\rightarrow 0$, the product yields $2\pi t \delta(x)$.

The reduced density matrix (\ref{redrho}) allows us to obtain the time evolution of the
neutrino populations and coherences, namely \be n_i(\vp,t) = \mathrm{Tr}\, \rho_r(t) a^\dagger_i(\vp)a_i(\vp) ~~;~~\mathcal{C}_{ij}(\vp,t) =  \mathrm{Tr} \, \rho_r(t) a^\dagger_i(\vp)a_j(\vp)~,~i\neq j \label{rhoel}\ee where the annihilation and creation
operators are in the Schroedinger picture. In the long time limit and using the results above, we find \be  n_1(\vp,t) = t ~  \Gamma_1(\vp) \cos^2\theta ~~;~~  n_2(\vp,t) = t ~ \Gamma_2(\vp) \sin^2\theta, \label{nuprod}\ee  where
           \be \Gamma_{1,2}  = \frac{2\pi\,g^2}{8 E^W_{\vk}} \int
           \frac{d^3\vQ}{(2\pi)^3\,E^e_{\vQ}\,\Omega_{1,2}}
        \delta\big( E^W_{\vk} -E^e_{\vQ}-\Omega_{1,2}\big)\label{partialwidths}\ee are the partial widths for the decay of the $W$ into the
            charged lepton and the neutrino mass eigenstates. For the off-diagonal matrix elements or coherences we find,
            \bea \mathcal{C}_{12}(\vp,t) = \mathcal{C}^\dagger_{21}(\vp,t)
            &=&
            \frac{2\pi\,g^2 \sin 2\theta}{32 E^W_{\vk}}~\frac{\sin[t\Delta]}{\Delta}~e^{i\Delta t} \nn \\
            && \int
           \frac{d^3\vQ}{(2\pi)^3\,E^e_{\vQ}\,\sqrt{\Omega_{1}\,\Omega_{2}}}\Big[ \delta\big( E^W_{\vk} -E^e_{\vQ}-\Omega_{1}\big)+
            \delta\big( E^W_{\vk} -E^e_{\vQ}-\Omega_{ 2}\big) \Big]. \nn \\
           \label{coher} \eea

           $dn_i(\vp,t)/dt$ yields the \emph{production  rate} of the neutrino
           mass eigenstates from the decay of the $W$ and the coherences $\mathcal{C}_{ij}$ are non-vanishing at any finite time. In Ref. \cite{glashow}, these coherences vanish
           as a consequence of the product of delta functions on the different mass shells
           of the mass eigenstates. The coherences vanish in the formal infinite time
           limit because of the oscillatory behavior averages out on time scales $t \gg 1/\Delta$ and they are of $\mathcal{O}\big(\sin(t\Delta)/t\Delta \big)$ with respect to the diagonal terms. However,
           we  emphasize that experimentally, the time
           scales involved (or   length scales) are of order $1/\Delta$ as these are
           the scales on which oscillatory phenomena are revealed.
           Taking $\Omega_1 \sim \Omega_2 \sim \oO$ in the
           denominators in (\ref{partialwidths},\ref{coher}), it
           follows that \be \mathcal{C}_{12}(\vp,t) \simeq \frac{\sin 2\theta}{2}\frac{\sin[t\Delta]}{\Delta}~e^{
           i\Delta t}
           ~\frac{1}{2}\big[\Gamma_1+\Gamma_2\big]. \label{cij}\ee
           Therefore, the coherences are of the same order of the
           population terms on time scales $t\leq 1/\Delta$, but
           average out for $t\gg 1/\Delta$,  showing that
           coherence persists over the oscillation time scale.

\subsection{Disentangling the neutrino: a two-time measurement}

As we have discussed above, a long baseline experiment is actually a \emph{two-time
measurement} process, as the charged lepton produced at the interaction vertex
at the \emph{source} is detected at the source or stopped and absorbed in a nearby ``beam dump''.

This
``measurement'' of the charged lepton disentangles the neutrinos from
the charged lepton in the quantum state
(\ref{finstate}) \cite{glashow}. The detection of the charged lepton
at the source (or its stopping at a nearby ``beam dump'')
projects the quantum state (\ref{finstate}) at
the observation time $t_S$ onto the single particle charged lepton
state $e^{-iE^e_{\vQ}t_S}~\Big|e_{\vQ}  \rangle     $ disentangling the
neutrino states into the ``collapsed'' state
\bea  \big|
\mathcal{V}_{e}(\vQ,t_S)\rangle & \equiv & \langle e_{\vQ}
\big|\Psi(t_S)\rangle ~e^{ iE^e_{\vQ}t_S}  \nn \\
&=& i\frac{g~e^{-iE_S
\frac{t_S}{2}}}{2\sqrt{2VE^W_{\vk} E^e_{\vQ}}}~
        \Bigg\{
    \frac{\sin\theta}{\sqrt{\Omega_{2, \vP}  }}~\big|\nu_{2,\vP}\rangle e^{ -i\Omega_{2,\vP} \frac{t_S}{2}}
    \left[\frac{\sin\Big[\big(E_S-\Omega_{2,\vP} \big)\frac{t_S}{2}\Big]}{    (E_S-\Omega_{2,\vP} )/2} \right]
    \nonumber \\ & & + \, \frac{\cos\theta}{\sqrt{ \Omega_{1,\vP} }}~\big|\nu_{1,\vP}\rangle
   e^{ -i\Omega_{1,\vP} \frac{t_S}{2}}
    \left[\frac{\sin\Big[\big(E_S-\Omega_{1,\vP} \big)\frac{t_S}{2}\Big]}{    (E_S-\Omega_{1,\vP} )/2} \right]
     \Bigg\}~; \label{colafinstate} \\
     && E_S = E^W_{\vk}-E^e_{\vQ}~;~\vP = \vk-\vQ. \nn \eea

      We note that up to a phase the
      coefficient functions that multiply $\big|\nu_{1,2}\rangle$
      are \be \frac{\sin\Big[(E_S-\Omega_1)\frac{t_S}{2}
      \Big]}{(E_S-\Omega_1)/2} ~~,~~\frac{\sin\Big[(E_S-\Omega_2)\frac{t_S}{2}
      \Big]}{(E_S-\Omega_2)/2}, \label{senos}\ee
      respectively. In the   limit $t_S
      \rightarrow \infty$, these
      become $2\pi \delta(E_S-\Omega_1)$ and $2\pi \delta(E_S-\Omega_2)$,
      respectively. Therefore, in this limit, for a fixed $E_S$, one of the quantum states will be projected out.
       However, as we insist on keeping a \emph{finite}
      time interval, we will keep $t_S$ finite.

      The state $\big|\mathcal{V}_{e}(\vQ,t_S)\rangle$ then evolves forward in time with the full Hamiltonian \be \big|\mathcal{V}_{e}(\vQ, t)\rangle = e^{-iH_0 t} U(t,t_S) e^{iH_0 t_S} \big|\mathcal{V}_{e}(\vQ,t_S)\rangle. \label{Nuoft}\ee The ``free'' evolution is obtained
     by setting  to lowest order $U(t,t_S)=1$, leading to \bea \big|\mathcal{V}_{e}(\vQ, t)\rangle & = &  i\frac{g~e^{-iE_S \frac{t_S}{2}}}{2\sqrt{2VE^W_{\vk} E^e_{\vQ}}}~
        \Bigg\{
    \frac{\sin\theta}{\sqrt{\Omega_{2, \vP}  }}~\big|\nu_{2,\vP}\rangle
    \Bigg[\frac{ \sin\Big[\big(E_S-\Omega_{2,\vP} \big)\frac{t_S}{2}\Big] }{
    (E_S-\Omega_{2,\vP} )/2} \Bigg]~e^{-i\Omega_{2,\vP}\frac{t_S}{2}}~e^{-i\Omega_{2,\vP}(t-t_S)}
    \nonumber \\ & + & \frac{\cos\theta}{\sqrt{ \Omega_{1,\vP}}}~\big|\nu_{1,\vP}\rangle
    \Bigg[\frac{\sin\Big[\big(E_S-\Omega_{1,\vP} \big)\frac{t_S}{2}\Big] }{ (E_S-
    \Omega_{1,\vP} )/2} \Bigg] e^{-i\Omega_{1,\vP}\frac{t_S}{2}}~e^{-i\Omega_{1,\vP}(t-t_S)}
     \Bigg\}. \label{timev}\eea
     The  phase factors $e^{-i\Omega_j t_S/2}$ multiplying each mass eigenstate are the consequence of the phase build-up during
    the time evolution from the production vertex until the detection of the charged lepton
    and the collapse of the wave function. These can be absorbed into the definition of the states $|\nu_{1,2}\rangle$ at the fixed time $t_S$.

     The expression (\ref{timev})   features the factors   \be \frac{\sin\Big[\big(E_S-\Omega_{j} \big)\frac{t_S}{2}\Big] }{ (E_S-
    \Omega_{j } )/2}, \label{dels}\ee
    which as ${t_S\rightarrow \infty}$ becomes $2\pi\, \delta(E_S-
    \Omega_{j } )$. These factors, which are a direct consequence of
     the neutrino state being produced by the decay of the ``parent'' particle (here the $W$) into an \emph{entangled}   quantum state, distinguish Eq. (\ref{timev}) from the familiar quantum mechanical description.
         These factors emerge from the (approximate) energy conservation at the decay vertex.
         Again, in the $t_S \rightarrow \infty$ limit, if the energy of the parent particle and the charged
         lepton are both certain, \emph{only one} of the mass eigenstates
         will be produced but not both.
         However, writing $\Omega_{1,2}$ as in Eq. (\ref{URO}), it follows that for
         $t_S \ll 4\oO/\delta m^2$, the width of the ``diffraction'' functions is \emph{much larger}
          than the frequency difference $\Delta$ and there is a substantial overlap between these ``approximate'' delta functions
           (see Fig. (\ref{fig:fpm})).
           Only for $t_S \geq t_{osc} = 2\pi/\Delta$ are the two peaks at $E_S -\oO = \pm \Delta$ actually resolved, whereas for $t_S\ll t_{osc}$, the two peaks are unresolved, ``blurred'' into one broad peak at $\oO$. Thus for $t_S \ll 2\pi/\Delta$, we can use the approximation \be \frac{\sin\Big[(E_S-\Omega_1)\frac{t_S}{2}
      \Big]}{(E_S-\Omega_1)/2} \simeq \frac{\sin\Big[(E_S-\Omega_2)\frac{t_S}{2}
      \Big]}{(E_S-\Omega_2)/2}\simeq \frac{\sin\Big[(E_S-\oO )\frac{t_S}{2}
      \Big]}{(E_S-\oO )/2}\,.
      \label{equalsenos}\ee

To illustrate the validity of the above approximation, let us
consider the case in which the typical size of the
source or beam dump is
a few meters across.
 In a typical experiment, the charged lepton
emerging from the interaction vertex travels this distance within a
time scale $t_S \approx 10^{-8}$ s, leading to an energy uncertainty above  $\delta E \sim \hbar/t_S \sim
10^{-7}\,\mathrm{eV}$. Taking as an example $\delta m^2 \sim
10^{-4}\,\mathrm{eV}^2$; $\oO \sim E_S \sim 100\,\mathrm{MeV}$, it
follows that $\delta m^2 /\oO \sim 10^{-12}\,\mathrm{eV} \ll \delta
E$.
Therefore, the detection or absorption of the charged lepton near the source,
 on a time scale much smaller than $1/\Delta$ cannot resolve the energy difference between the mass eigenstates and the approximation (\ref{equalsenos}) is justified.

Another approximation we can use in (\ref{colafinstate}) is $\Omega_1 \simeq \Omega_2\simeq \oO$,  for $\oO \gg \Delta$. Absorbing the phase factors $e^{-i\Omega_j t_S/2}$
       into the definition of the states $|\nu_j\rangle$,   the time evolved disentangled state is then approximately given by \be   \big|
\mathcal{V}_{e}(\vQ,t)\rangle    \simeq     \frac{i g }{ \big[8
VE^W_{\vk} E^e_{\vQ}\oO\big]^\frac{1}{2}}
\left[\frac{\sin\Big[\big(E_S-\oO \big)\frac{t_S}{2}\Big]}{
(E_S-\oO  )/2} \right]~
        \Bigg\{     {\sin\theta} ~\big|\nu_{2,\vP}\rangle~e^{-i\Omega_{2,\vP}(t-t_S)}
     + \, {\cos\theta} ~\big|\nu_{1,\vP}\rangle~e^{-i\Omega_{1,\vP}(t-t_S)}
    \Bigg\},  \label{colafinstatesmallt}
   \ee
   for $t_S \ll 2\pi/\Delta$.

   The state inside the brackets is identified as the usual quantum mechanical state that is
   time evolved from the ``electron'' neutrino state, which is prepared initially at $t_S$.
    From this analysis, we see that there are two conditions required for the disentangled neutrino
     state to be identified with the familiar quantum mechanical state: $\delta m^2/\oO^{\,2} \ll1$ and $t_S \ll t_{osc} \sim 2\pi \oO/\delta m^2$. The former is always satisfied for neutrinos with
     $\delta m^2 \sim 10^{-3}-10^{-4} \,\mathrm{eV}^2,\ \oO > \,\mathrm{few~MeV}$, while
      the latter is fulfilled
      if the charged lepton produced at the source is either observed or stopped at a beam-dump near the production region
       in long-baseline experiments.

       The latter condition implies that
       the neutrino state is \emph{disentangled  before oscillations can
       occur}. In a long-baseline experiment this is achieved if the charged lepton, which is entangled with the neutrino at the production
        vertex, is measured or stopped near the production region. Therefore, we conclude that an important condition for the familiar quantum mechanical description of oscillations to be reliable is that the quantum state must be disentangled on time scales \emph{much shorter} than the oscillation time.

        It is clear from this discussion that the precise value of $t_S$ is \emph{irrelevant} as long as $t_S \ll t_{osc}$. Experimentally, $t_S$ is of the order of the time of
        flight of the charged lepton in the production region at the source namely a few $10^{-8}\,s$.

\subsection{Transition amplitudes and event rates}

The number of charged lepton events with momentum $\vQ$ produced at the source, at time $t_S$ is given by \be  n_e(\vQ,t_S)= \langle
\Psi_e(t_S)| a^\dagger_{e}(\vQ) a_e(\vQ) | \Psi_e(t_S) \rangle
=\langle \mathcal{V}_{e}(\vQ,t_S) \big|
\mathcal{V}_{e}(\vQ,t_S)\rangle.  \label{neND} \ee For $t_S \ll
t_{osc}=2\pi/\Delta$ and $\oO \gg \Delta$, using the approximations
leading  to (\ref{colafinstatesmallt}), we obtain the charged lepton differential
detection rate
 at the source

\bea
(2\pi)^3\, \frac{dN^{S}_e}{d^4x\, d^3\vQ}&=&\frac{dn_e(\vQ,t_S)}{dt_S} \nn \\
&=&  \frac{2 g^2 }{ 8 V E^W_{\vk} E^e_{\vQ}\, \oO  } \,\frac{\sin\Big[\big(E_S-\oO
\big) {t_S }\Big] }{    (E_S-\oO  ) }  \nn \\
 &\simeq&  \frac{ 2\pi g^2 }{ 8
VE^W_{\vk} E^e_{\vQ }\,\oO  }~ \delta(E_S-\oO ), \label{diffrateND}\eea
where at the last step we have replaced the diffraction function by the delta function.  This can be justified as follows. For $t_S \sim 10^{-8} \,\mathrm{s}$, the width of this function (the resolution)  in energy is $\sim 10^{-7}\,\mathrm{eV}$. Since the typical energy in a long-baseline experiment is $ \gtrsim 40-100 \,\mathrm{MeV}$, the error incurred in replacing the diffraction function by a delta function is smaller than one part in $10^{15}$.


     We can also obtain the transition amplitude for the disentangled  state to produce
     a final charged lepton and another $W$ particle at the far detector at time $t_D$,
     where $t_D-t_S \sim L$ and $L$ is the baseline. This is obtained from the time evolution of the disentangled state with the full Hamiltonian from the time of disentanglement $t_S$ until the time $t_D$, at which the detection measurement is carried out by overlapping the time evolved state with the final state.

      The transition amplitude is given by\bea \mathcal{A}_{\alpha \rightarrow \beta} &=& \langle W(\vk_D),l_\beta(\vp_D)\big|
     \,e^{-iH(t_D-t_S)}\,\big|\mathcal{V}_{e}(\vQ,t_S)\rangle  \nn \\
     &=& e^{-iE_D t_D}\, \langle W(\vk_D),l_\beta(\vp_D)\big|\,U(t_D,t_S)\,e^{iH_0t_S}\,\big|\mathcal{V}_{e}(\vQ,t_S)\rangle. \label{Aabcola}\eea
     The disappearance and appearance amplitudes are then given by
     \bea \mathcal{A}_{e\rightarrow e} & = & -g^2\,\Pi\,(2\pi)^3 \delta(\vk_S-\vp_S-\vk_D-\vp_D)  \times \nn \\
     & & \Bigg\{ \frac{\cos^2\theta}{2\Omega_{1,\vP}} e^{-i\Omega_{1,\vP}\frac{t_D}{2}}\, \Bigg[\frac{\sin\Big[\big(E_S-\Omega_{1,\vP} \big)\frac{t_S}{2}\Big] }{ (E_S-
    \Omega_{1,\vP} )/2} \Bigg] \, \Bigg[\frac{\sin\Big[\big(E_D-\Omega_{1,\vP} \big)\big(t_D-t_S\big)/2\Big] }{ (E_D-\Omega_{1,\vP} )/2} \Bigg]
    \nonumber \\ &  & ~~+ \frac{\sin^2\theta}{2\Omega_{2,\vP}} e^{-i\Omega_{2,\vP} \frac{t_D}{2}}\,
     \Bigg[\frac{\sin\Big[\big(E_S-\Omega_{2,\vP} \big)\frac{t_S}{2}\Big] }{ (E_S-
    \Omega_{2,\vP} )/2} \Bigg] \, \Bigg[\frac{\sin\Big[\big(E_D-\Omega_{2,\vP} \big)\big(t_D-t_S\big)/2\Big] }{ (E_D-
    \Omega_{2,\vP} )/2} \Bigg]\Bigg\} \nn \\
    \label{disadisent}\eea
and
     \bea \mathcal{A}_{e\rightarrow \mu} & = & -g^2\,\Pi\,(2\pi)^3 \delta(\vk_S-\vp_S-\vk_D-\vp_D) \,\frac{\sin2\theta}{2} \times \nn \\
     && \Bigg\{ \frac{e^{-i\Omega_{1,\vP}\frac{t_D}{2}}}{2\Omega_{1,\vP}}  \, \Bigg[\frac{\sin\Big[\big(E_S-\Omega_{1,\vP} \big)
     \frac{t_S}{2}\Big] }{ (E_S-\Omega_{1,\vP} )/2} \Bigg] \,
      \Bigg[\frac{\sin\Big[\big(E_D-\Omega_{1,\vP} \big)\big(t_D-t_S\big)/2\Big] }{ (E_D-\Omega_{1,\vP} )/2} \Bigg]
    \nonumber \\ &   & ~~  - \frac{e^{-i\Omega_{2,\vP} \frac{t_D}{2}}}{2\Omega_{2,\vP}}
      \, \Bigg[\frac{\sin\Big[\big(E_S-\Omega_{2,\vP} \big)\frac{t_S}{2}\Big] }{ (E_S-
    \Omega_{2,\vP} )/2} \Bigg] \,
     \Bigg[\frac{\sin\Big[\big(E_D-\Omega_{2,\vP} \big)\big(t_D-t_S\big)/2\Big] }{ (E_D-
    \Omega_{2,\vP} )/2} \Bigg]\Bigg\},\nn \\
    \label{aperadisent}\eea
with $\vP=\vk_S-\vp_S$, and $\Pi$ is given by \be \Pi =
\Big[\frac{1}{16~V^4~E^W_{\vk_S}\,E^W_{\vk_D}\,E^l_{\vp_S}\,E^l_{\vp_D}}
\Big]^\frac{1}{2} ~~\,, \label{Pifactor}\ee   the labels $D,S$ refer to (far) detector and source respectively.

Implementing the same approximations leading to the factorized state
(\ref{colafinstatesmallt}), namely $\oO \gg \Delta$ and $\,t_S\Delta \ll
1$, we find the disappearance and appearance probabilities
\bea \mathcal{P}_{e\rightarrow e} (t_D) & = &  \Big(\frac{g^2 \Pi}{2\oO_{\vP}}\Big)^2 (2\pi)^3 \, V\,\delta(\vk_S-\vp_S-\vk_D-\vp_D)\,   2\pi\,t_S\,\delta(E_S-\oO_{\vP}) \nonumber \\ && \Bigg\{\cos^4\theta \, f^2_+(x,t,\Delta) + \sin^4\theta \,f^2_-(x,t,\Delta)+ 2\cos^2\theta \sin^2\theta \cos(t\Delta) f_+(x,t,\Delta)f_-(x,t,\Delta) \Bigg\}, \label{probee}\nonumber \\
 \mathcal{P}_{e\rightarrow \mu}(t_D)  & = &  \Big(\frac{g^2 \Pi}{2\oO_{\vP}}\Big)^2 (2\pi)^3 \, V\,\delta(\vk_S-\vp_S-\vk_D-\vp_D)\,   2\pi\,t_S\,\delta(E_S-\oO_{\vP})\, \frac{\sin^22\theta}{4}\nonumber \\ && \Bigg\{f^2_+(x,t,\Delta)+f^2_-(x,t,\Delta)-2\cos(t\Delta) f_+(x,t,\Delta)f_-(x,t,\Delta) \Bigg\}, \label{probemu}\eea
where $t= t_D-t_S$ and $x=E_D-\oO_{\vP}$. Here,
\be   E_S = E^W_{\vk_S}-E^e_{\vp_S}~~;~~ E_D= E^W_{\vk_D}+E^l_{\vp_D}\label{defs2},\ee
and $f_{\pm}$ are given by Eq. \ref{fpm}.

In the long time limit, using $f_{\pm}(x,t,\Delta) \rightarrow 2\pi \,t\,\delta(x\pm\Delta)$ their product is given by (\ref{ltff}), and we find
\bea \mathcal{P}_{e\rightarrow e}(t_D)  & = &  \Big(\frac{g^2 \Pi}{2\oO_{\vP}}\Big)^2 (2\pi)^5 \, V\,\delta(\vk_S-\vp_S-\vk_D-\vp_D)\,
\,t_S\,\delta(E_S-\oO_{\vP}) \nonumber \\ && \Bigg\{\cos^4\theta \, t\,\delta(x+\Delta) + \sin^4\theta \,t\,\delta(x-\Delta)+ 2\cos^2\theta
\sin^2\theta \,\frac{\sin(2\,t\,\Delta)}{2\Delta}\,\frac{1}{2}\Big[\delta(x+\Delta)+\delta(x-\Delta)\Big] \Bigg\}, \nonumber \\
 \mathcal{P}_{e\rightarrow \mu}(t_D)  & = &  \Big(\frac{g^2 \Pi}{2\oO_{\vP}}\Big)^2 (2\pi)^5 \, V\,\delta(\vk_S-\vp_S-\vk_D-\vp_D)\,
  \,t_S\,\delta(E_S-\oO_{\vP})\, \frac{\sin^22\theta}{4}\nonumber \\ && \Bigg\{t\,\delta(x+\Delta)+t\,\delta(x-\Delta)-2\,\frac{\sin(2\,t\,\Delta)}{2\Delta}\,\frac{1}{2} \Big[\delta(x+\Delta)+\delta(x-\Delta)\Big]  \Bigg\}. \label{probemult}\eea where $\vec{P}=\vec{k}_D+\vec{p}_D$.

 These transition probabilities  feature the \emph{two} time scales $t_S$ and $t=t_D-t_S$ associated with the measurements at the near and far detector. They also feature energy conserving delta functions associated with the different mass eigenstates.

 There is a further simplification when $\oO \gg \Delta$. In this regime, when the probabilities (\ref{probemult}) are \emph{integrated  over a smooth density of states}, the delta functions corresponding to the mass eigenstates yield the density of states at values $E_D = \oO \mp \Delta$. In typical experiments, where $\oO \sim 100 \,\mathrm{MeV}$ and $\delta m^2 \sim 10^{-3}\,\mathrm{eV}^2$, the density of final states must vary dramatically
 near $\oO$ to resolve the small interval $\Delta$, with $\Delta/\oO \lesssim 10^{-19}$. Therefore,
  understanding the probabilities as being integrated with a smooth final density of states insensitive to the mass difference,
   we can approximate $\delta(x\pm \Delta) \simeq  \delta(x)$. In this case, we can approximate the above expressions by
 \bea \mathcal{P}_{e\rightarrow e} (t_D) & = &  \Big(\frac{g^2 \Pi}{2\oO_{\vP}}\Big)^2 (2\pi)^5 \, V\,\delta(\vk_S-\vp_S-\vk_D-\vp_D)\,
   \,t_S\,\delta(E_S-\oO_{\vP})\,\delta(E_D-\oO_{\vP}) \nonumber \\ && \Bigg\{t\Big[\cos^4\theta   + \sin^4\theta \Big]+
   2\cos^2\theta \sin^2\theta \,\frac{\sin(2\,t\,\Delta)}{2\Delta}  \Bigg\}, \label{probeelt2} \\
 \mathcal{P}_{e\rightarrow \mu} (t_D) & = &  \Big(\frac{g^2 \Pi}{2\oO_{\vP}}\Big)^2 (2\pi)^5 \, V\,\delta(\vk_S-\vp_S-\vk_D-\vp_D)\,
   \,t_S\,\delta(E_S-\oO_{\vP})\,\delta(E_D-\oO_{\vP})\, \frac{\sin^22\theta}{2}\nonumber \\ && \Bigg\{t - \,\frac{\sin(2\,t\,\Delta)}{2\Delta} \Bigg\}.
    \label{probemult2}\eea The product $ \delta(E_S-\oO_{\vP})\,\delta(E_D-\oO_{\vP}) $ is an \emph{approximate} energy conservation at both production and detection vertices, where we have neglected $\Delta$, which is twice the energy difference between the mass eigenstates.

 Further insight can be gained by obtaining the phase space distribution of the number of charged leptons $l=e,\mu$ at the far detector
 \be (2\pi)^3\,\frac{dN^{FD}_l}{d^3x \,d^3\vp_D}=n_l(\vp_D,t_D) = \langle \mathcal{V}_e(\vQ,t_D,t_S)\big|a^\dagger_l(\vp_D)a_l(\vp_D) \big| \mathcal{V}_e(\vQ,t_D,t_S)\rangle. \label{fardetn}\ee
 Here, \be \big| \mathcal{V}_e(\vQ,t_D,t_S)\rangle = e^{-iH_0 t_D}U(t_D,t_S)e^{iH_0t_S} \big| \mathcal{V}_e(\vQ,t_S)\rangle \label{tevolnu}\ee
  is the neutrino state disentangled at $t_S$
  near the source and has been time-evolved until it is detected
   at the far detector at time $t_D$. Not surprisingly, since we only keep terms that are up to order $g^2$,
   the time evolved state contains a single lepton Fock state. We find that  \be (2\pi)^3\,
   \frac{dN^{FD}_e}{d^3x \, d^3\vp_D} = \mathcal{P}_{e\rightarrow e} (t_D) ~~;~~ (2\pi)^3\, \frac{dN^{FD}_{\mu}}{d^3x \,d^3\vp_D} =
    \mathcal{P}_{e\rightarrow \mu} (t_D), \label{nemuFD}\ee with the probabilities $\mathcal{P}_{e\rightarrow e} (t_D)$
    and $\mathcal{P}_{e\rightarrow \mu} (t_D)$ are given by (\ref{probeelt2},\ref{probemult2}).

 Taking the \emph{time derivative} with respect to $t_D$, we obtain the differential charged lepton  event \emph{rates} at the far detector  \bea (2\pi)^3\frac{dN^{FD}_e}{d^3x \,dt \,d^3\vp_D}  & = &   \Big(\frac{g^2 \Pi}{2\oO_{\vP}}\Big)^2 (2\pi)^5 \, V\,\delta(\vk_S-\vp_S-\vk_D-\vp_D)\,   \,t_S\,\delta(E_S-\oO_{\vP})\,\delta(E_D-\oO_{\vP}) \nonumber \\ && \Bigg\{ \cos^4\theta   + \sin^4\theta  + 2\cos^2\theta \sin^2\theta \, \cos(2t\Delta)  \Bigg\} ,\label{FDee}\eea
 \bea  (2\pi)^3\frac{dN^{FD}_{\mu}}{d^3x \,dt\, d^3\vp_D} & = &
 \Big(\frac{g^2 \Pi}{2\oO_{\vP}}\Big)^2 (2\pi)^5 \, V\,\delta(\vk_S-\vp_S-\vk_D-\vp_D)\,
   \,t_S\,\delta(E_S-\oO_{\vP})\,\delta(E_D-\oO_{\vP})\, \frac{\sin^22\theta}{2}\nonumber \\ &&
   \Bigg\{1 -  \cos(2t\Delta) \Bigg\}. \label{FDemu}\eea Remarkably, these rates can be simply
   written as \be  (2\pi)^3\frac{dN^{FD}_\beta}{d^3x \,dt\, d^3\vp_D} =
   (2\pi)^3\frac{dN^{S}_\alpha}{d^3x \,d^3\vp_s}\,P_{\alpha \rightarrow \beta}(t) \,d\Gamma_{\nu_\beta \rightarrow W\,l_\beta}, \label{factorized}\ee
   where we have used the expression (\ref{diffrateND}) for
    the differential charged lepton event rate at the
   source and integrated in
    $t_S$,
    \be d\Gamma_{\nu_\beta \rightarrow W\,l_\beta} = \frac{(2\pi)^4\,g^2\,V}{8V^3E^W_{\vk_D}E^{l^\beta}_{\vp_D}\oO_{\vP}}
     \, \delta(\vk_S-\vp_S-\vk_D-\vp_D)\,\delta(E_D-\oO_{\vP}) \label{prodrateFD}\ee is the charged lepton production
     rate from the reaction $\nu_\beta \rightarrow W\,l_{\beta}$ for a \emph{flavor} neutrino of energy $\oO$ and
     $P_{\alpha \rightarrow \beta}(t)$ are the disappearance ($\alpha = \beta$) or appearance ($\alpha \ne \beta$) transition probabilities obtained from the usual \emph{quantum mechanical} calculations of Rabi-oscillations, \bea P_{ e \rightarrow  e} & = &  1-\sin^2(2\theta)\,\sin^2\Big[\frac{\delta m^2}{4\oO}\,t \Big] \,,\nonumber \\ P_{ e \rightarrow \mu} & = &  \sin^2(2\theta)\,\sin^2\Big[\frac{\delta m^2}{4\oO}\,t \Big] \,. \eea

 The remarkable aspect of the final result (\ref{prodrateFD}) is the \emph{factorization}
 of the different processes contributing to the far detector event \emph{rate}, namely the number
 of events at the
  source multiplies the quantum mechanical transition probability which in
 turn multiplies the production rate at the vertex in the far detector. This factorization is a
 distinct consequence of the \emph{two time analysis}, of the disentanglement of the neutrino
  near the production region
   along with the approximations invoked in the resolution of the energy
  conserving delta functions. We emphasize that the factorization in terms of the
 quantum mechanical transition probabilities $P_{\alpha \rightarrow \beta}(t)$ \emph{only} applies to the detection \emph{rate}
  defined by taking the time derivative. The \emph{total number of events} per phase space volume also factorizes but \emph{not} in terms of the quantum mechanical transition probabilities but in terms of their integral in time.

\subsection{Possible corrections for short baseline experiments}
 This analysis also suggests that potentially important
        corrections may arise in \emph{short baseline}
        experiments for much lower energy. In this case, it may occur that the time scale of
        disentanglement is of the same order as the oscillation
        scale $\oO/\delta m^2$ so that $t_S \delta m^2/\oO \sim 1$ and
        the sine functions cannot be factored out of the quantum
        state as in (\ref{colafinstatesmallt}).  In this case, the resulting disentangled state is  \be   \big|
\mathcal{V}_{e}(\vQ,t)\rangle    \propto
        \Bigg\{     {\sin\theta} ~\big|\nu_{2,\vP}\rangle~e^{-i\Omega_{2,\vP}(t-t_S)}\,f_-(x,t_S,\Delta)
     + \, {\cos\theta} ~\big|\nu_{1,\vP}\rangle~e^{-i\Omega_{1,\vP}(t-t_S)}\,f_+(x,t_S,\Delta)
    \Bigg\},  \label{nodis}
   \ee
                which is \emph{not} of the
        familiar quantum mechanical form of a coherent superposition
        of the mass eigenstates with the usual
        $\cos\theta,\sin\theta$ amplitudes but with the ``diffraction
        functions'' multiplying these factors.  If the width of these functions is of order $2\pi/t_S \lesssim 2\Delta$, interference term in the probability will feature
        the product (\ref{ltff}) which would lead to an extra interference terms of the
        form  $\sin(t_S\Delta)/t_S\Delta$ multiplying the $\sin(2t\Delta)/2\Delta$ in
         (\ref{probeelt2},\ref{probemult2}). Such factor yields an \emph{extra modulation with
         energy} which may yield phenomenologically interesting modifications in the interpretation and analysis of data.
         Although the typical time scale $t_S$ for disentanglement near the source region is approximately the same for short and long baseline experiments, this discussion applies to the possibility  in which $t_S$ is not too small compared
         to the oscillation time scale as \emph{could} be the case in short-baseline experiments. For example at LSND the source region is a few meters across and the baseline is $\sim 30 \,m$, and at KARMEN where the baseline is $\sim 17.5\,m$. The baseline and beam energy are designed so that at least an oscillation takes place along the baseline. However in these short-baseline experiments, the distance over which the neutrino state is disentangled from the charged lepton, a few meters, is not too much smaller than the baseline, consequently $t_S$ may not be much smaller than an oscillation time.

         The potential phenomenological importance of these effects merit their further study.

\subsection{Wave packet description}

  As mentioned earlier in the paper, we restricted ourselves to plane-wave states, the main reason being to discuss the main concepts within the simplest setting. Although we postpone a thorough discussion of the more technical aspects including the wave packet description to further study, we can provide physical arguments that allow the extrapolation of the main results obtained above to the wave packet case.

  Consider that the initial particle, the charged leptons measured at the near
   and far detectors and the final particle are all described by wave packets with
  a \emph{macroscopic} localization length scale $\sigma$ of the order of the typical scale of the
    detectors (a few meters) localized at the source and detectors respectively. These are sharply localized in momentum. The transition amplitudes between wave packets are now obtained from (\ref{Aabcola}-\ref{aperadisent}) by convolution with the wave functions of the initial state, the charged lepton measured
    near the source and those of the final state at the far detector.

     The disentangled neutrino state (\ref{colafinstate})
 is now   described as a propagating wave packet with a typical
  spatial localization length also of $\mathcal{O}(\sigma)$, namely of macroscopic scale. These mildly localized
    wave packets lead to both momentum and energy uncertainty of the
    same order $\Delta E  \sim \Delta p  \sim \hbar/\sigma \sim
    \hbar/t_S \sim 10^{-7}\,\mathrm{eV}$. These uncertainties in
    energy and momentum are \emph{much larger  than typical values
    for the neutrino energy differences} $\delta m^2/\oO \sim
    10^{-12}\,\mathrm{eV}$. As this wave packet propagates,
  the mass eigenstates will slowly separate because they feature different group velocities, and coherence will be maintained
 as long as the separation between the wave
  packets is much smaller than their localization lengths (neglecting
  dispersion). For a baseline $L$ and typical energies $\oO \sim 100\,\mathrm{MeV}$, the separation at the far detector is $\propto
  (\Delta/\oO) ~L \sim 10^{-20} L$. Therefore, coherence is always maintained for macroscopic localization lengths in terrestrial experiments.
   The detection amplitudes (\ref{disadisent},\ref{aperadisent}) and probabilities (\ref{probeelt2},\ref{probemult2}) vanish  until the time when the front of the packet overlaps with the far detector, namely for $t_D-t_S \approx L$,
   with an  uncertainty   of $\mathcal{O}(\sigma)$, which for long-baseline
  experiments is always $\sigma \ll L$. The total number of events at the far detector is
  found by integrating the rate (\ref{prodrateFD}) from $t = L$, which is when the front of the wave packet overlaps with the detector  until $t = L
  +\sigma$ when the wave packet has completely passed through the detector. For $L \gg \sigma$ and neglecting the separation of wave packets the total number of events at the far detector is obtained by replacing (\ref{factorized}) by

  \be   \frac{dN^{FD}_\beta}{d^3x  \, d^3\vp_D} =
    \frac{dN^{S}_\alpha}{d^3x \,d^3\vp_s}\,\left[P_{\alpha \rightarrow \beta}(L)\,\sigma\right]
    \,d\Gamma_{\nu_\beta \rightarrow W\,l_\beta}\,.
    \label{factorizedwp}\ee The delta functions will also be
    integrated with the wave packet profiles and volume factors are replaced by $\sigma^3$. Furthermore, at
    distances from the source $\gg \sigma$, the disentangled neutrino wave packet
    propagates as a spherical wave. Therefore, for far-detectors in
    the ``radiation zone,'' the probability will be suppressed by a
    geometric flux factor $1/L^2$ \cite{stock,dolgov}.
    Obviously, there are uncertainties of $\mathcal{O}(\sigma)$ in this argument, but
    these
  are irrelevant for long-baselines $L \gg \sigma$. The $\sigma$ multiplying the total number of events in (\ref{factorizedwp}) is expected: it is the ``total interaction time'' during which the wave packet interacts with the detector. Obviously, these factors modify the overall normalization. However, the distortion of the energy spectrum in disappearance and the appearance probabilities for long-baseline experiments are determined by $P_{\alpha \rightarrow \beta}(L,E)$ (for short baseline there are potential corrections as discussed above). While we deem
    these arguments describing the wave-packet
  scenario to be physically sound,   we will provide a thorough technical discussion of this   case based on the results obtained above in the plane wave
  approximation in future work. The analysis presented above for the dynamics of disentanglement and propagation provides the fundamental basis to include the wave packet description, which is obtained from simple convolution of these results with
  wave packet wave functions.

\section{Conclusions and further discussion}
In appearance and disappearance   experiments, neutrinos produced by
a charged current vertex at the source   are \emph{entangled} with the charged lepton. We study the \emph{dynamics} of this quantum state directly as a function of time.

       If the charged
       lepton (or daughter particle) produced at the source is not measured, tracing
       out this degree of freedom yields a density matrix for the
       neutrino. In the mass basis, the time evolution of the diagonal density matrix elements describe
       the production of mass eigenstates from the decay, whereas the off-diagonal density matrix
       elements exhibit the oscillations resulting from the
       interference of mass eigenstates and
        are a measure of \emph{coherence}. We find that coherence remains up to a time of the order
        of the oscillation time scale during which diagonal and off diagonal matrix elements are of the same order.

     The \emph{measurement}   of the charged lepton
     near the source (or its stopping at a nearby ``beam dump'')
     \emph{disentangles} the neutrino
       state, and it is the further time evolution of this
       disentangled state with the total Hamiltonian that
       leads to the production of charged leptons at the far
       detector. Thus, the process of production and detection in
       long-baseline experiments   involves \emph{two different time scales}: the measurement
       of the charged lepton near the source determines the first time scale at which the neutrino
       state is \emph{disentangled}, while the measurement of the charged lepton at the far detector is the second
         time scale, which is  much longer than the first one in long-baseline experiments.

         In this time dependent description, we establish that the usual quantum mechanical
         state emerges if the disentanglement  of the charged lepton produced with the neutrino occurs
         on time scales \emph{much} shorter than the oscillation
         scale $\sim E/\delta m^2$. Under these circumstances, the
           event rate at the far detector factorizes in terms of the usual quantum mechanical probability if the
         final density of states is insensitive to the difference in
         the energies of the mass eigenstates in the
         ultrarelativistic limit $\delta m^2/E$.

         Although in this article we focused on the study of the dynamics for the case of plane
         waves to exhibit the main results within the simplest
         setting, we provide physically motivated arguments to
         extrapolate the results to the case of wave packets, analyzing the
         effect of localization of the initial and final state wave functions at the near and far detector on the detection process. Combining the results obtained for plane waves with a wave
         packet description yields the total number of events
         detected in terms of the usual transition probability as a
         function of baseline. A deeper and more detailed analysis of wave packets and localization aspects    will be provided in forthcoming work.

         The analysis presented here also suggests that  there \emph{could}
         arise potentially interesting corrections in the case of
         \emph{short baseline} experiments such as MiniBoone  and/or LSND with baselines of $\sim 500,\sim 30$ meters and typical energies $\sim 1\mathrm{GeV};\sim 30\mathrm{MeV}$, respectively,  wherein the
         disentanglement of the neutrino could occur on time scales
         of the \emph{same order} as the oscillation scale. In this
         case, the disentangled state will differ from the usual
         quantum mechanical coherent superposition and this
         difference introduces \emph{extra modulation with energy} and could be important in the final detection rate.
         Such   possibility could be relevant in the interpretation and analysis of
         data, in particular masses and mixing angles and certainly merits further detailed study.

\acknowledgements
The authors thank Evgeny Akhmedov, Hamish Robertson and  Alexei Smirnov for fruitful correspondence and comments. DB and JW are supported by NSF grants: PHY-0553418
and PHY-0852497. JW thanks support through a Daniels and Mellon Fellowships. R. H. and J. H. are supported by the DOE through Grant No. DE-FG03-91-ER40682.


\end{document}